\documentclass[12pt,leqno]{article}
\usepackage{amsmath,amsthm,amssymb}

\def\init{\setcounter{equation}{0}}
\newtheorem{theorem}{Theorem}[section]

\newcommand{\Z}{\mathbb{Z}}

\newcommand{\e}{{\varepsilon}}

\newcommand{\rw}{\rightarrow}

\usepackage{tikz}
\usetikzlibrary{decorations.pathreplacing}

\usepackage{verbatim}

\title{Hawking radiation from  not-extremal  and extremal  Reissner-Nordstrom  black holes}

\author{G.Eskin, \ \ \  Department of Mathematics, UCLA,\\ Los Angeles,
CA 90095-1555, USA. \ E-mail: eskin@math.ucla.edu
}

\begin{document}

\maketitle

\begin{flushright}
\Large To dear Shmuel Agmon  with great  admiration
\\
\
\\
\end{flushright}

\begin{abstract}
We consider  the non-extremal  Reissner-Nordstrom  black  hole  and construct a wave packet that exhibits the Hawking radiation.
We    find  the average  of the number  of the created particles  with respect  to the $|0\rangle$ vacuum  state  and with respect  to Unruh type  vacuum state.
The  average  of the number operator  in the $|0\rangle$ vacuum state  consists  of two terms:  one is related to the Hawking radiation  and the second  is
not related.  We  use the  same construction  for the extremal RN black  hole  and get  that  the average  of the number  operator with respect
  to the $|0\rangle$  vacuum state  is also a sum  of two  term,  where the one related  to the Hawking radiation  is equal to zero.
  
  This result  is consistent  with  other  works  on the Hawking radiation for  the extremal RN black hole.
 
\end{abstract}

\section{Introduction}
\init
The phenomenon  that the black hole emits quantum  particles  was discovered by S.Hawking  in the seminal paper [7]
and it  is called  the Hawking radiation.  Any wave  packet  with the initial conditions  having the  support  outside
the black hole  produces  some particles  emission.  The Hawking  radiation refers  to the creation of particles intimately 
related  to the black  hole.

In [8]
to get  the Hawking radiation   one takes  the limit when  the time tends  to $-\infty$.  A rigorous 
derivation of Hawking radiation by this way was obtained by Fredenhagen  and Haag in [6]  for the case  of Schwartzschield  black  hole.

In [3]  we developed  a new  approach  to study the Hawking  radiation   in the case of a rotating black hole with a variable velocity.  Note that 
the previous methods  of [7], [8]  and others  do not work in this case.  We shall use  the approach  of paper [3]  in the present paper too.
 
  We consider  the case  
of two  space dimensions  and  we construct  a special  wave  packet of  the form
$$
(\rho-r_+)^\e e^{-a(\rho-r_+)}e^{iS(x_0,\rho,\varphi)} \ \ \mbox{at}\ \ x_0=0,
$$
where  $(\rho,\varphi)$  are  polar  coordinates,  $\rho>r_+,\ \ r_+$  is the event horizon,  $S(x_0,\rho,\varphi)$  is an approximate  eikonal  function
that tends  to the  infinity  when  $\rho\rw r_+$.
Here constant $a$ plays an important role   since  $\mbox{supp}\,(\rho-r_+)^\e e^{-a(\rho-r_+)}$  tends  to  $\rho=r_+$  when $a\rw\infty$.

Normalizing the wave  packet,  considering  the average  of the number  of particles  operator  and taking the limit as $a\rw\infty$,
we  get the the Hawking radiation.

Thus,  taking the limit  as $a\rw\infty$  is replacing  the necessity  of taking  the limit  $T\rw -\infty$   as in  Fredenhagen-Haag approach.

There are  two choices  of selecting  the vacuum state: 1)  the vacuum state  $|0\rangle$  that  was used
in [7]  and  [8]  and  others,  and 2)  the  Unruh type   vacuum  state  that was used in [3],  which is similar to the original Unruh 
vacuum state [10].  Note that  the limit  
when $a\rw\infty$  
of     the average of the number operator with respect  to 
$|0\rangle$ state
 is  the sum  of two terms  $I_+$  and  $I_-$,  where  $I_+$  corresponds  to the Hawking 
 radiation and $I_-$  is not  
(cf. \S 3).

Note  that in the case of  the average  of the number of particles with respect   to  the  Unruh type  vacuum 
all terms correspond  to the  Hawking  radiation  (cf. \S4).

In \S  5  we consider  the case  of extremal RN black hole.  
Our approach allows an uniform treatment  of extremal and non-extremal  RN black holes.
It is shown  that the  limit  when $a\rw \infty$  of the  average  of the  number
operator  with  respect  to $|0\rangle$ vacuum  also  contains  two terms  $I_++I_-$  
and the one  corresponding to the Hawking  radiation is zero. 
 This is consistent  with the conclusion  of other  
authors ([1],  [2],  [9])  that presents  different arguments  to  conclude  the Hawking radiation  for the  extremal  RN  black  hole  is zero.

The  plan  of the paper  is the following:

In \S2  we present  the elements  of the second  quantization  as  in the paper  [3] following  mostly  the lecture notes  of Jacobson  [8].

In \S 3  we study  not-extremal RN  black holes  when the vacuum state  is the $|0\rangle$ vacuum.

In \S4 we consider  not-extremal RN  black  holes  when the vacuum state is  the Unruh type vacuum state.

In \S5  we consider  the extremal RN black holes     when the vacuum states are
the $|0\rangle$ vacuum  and the Unruh type  vacuum.

\section{Elements  of the quantum  field  theory}
\init

The Reissner-Nordstrom  metric  in the  case  of two space  dimensions  has  the following  form  in polar  coordinates $(\rho,\varphi)$  (cf.  [4]):
\begin{multline}																				\label{eq:2.1}
ds^2=\Big(1-\frac{2 m}{\rho}+\frac{e^2}{\rho^2}\Big)dx_0^2-d\rho^2-\rho^2\,d\varphi^2-2\Big(\frac{2 m}{\rho}-\frac{e^2}{\rho^2}\Big)dx_0d\rho -
\Big(\frac{2 m}{\rho}-\frac{e^2}{\rho^2}\Big)d\rho^2,
\end{multline}
where $ m$  is  the mass  and  $e$  is  the electrical  charge  of the black hole.  The corresponding  wave equation $\Box_g u=0$  has  the form  
\begin{multline}																				\label{eq:2.2}
\frac{\partial^2 u}{\partial x_0^2}-\frac{\partial^2 u}{\partial \rho^2}-\frac{1}{\rho^2}\frac{\partial^2 u}{\partial\varphi^2}-\Big(-\frac{\partial}{\partial x_0}+\frac{\partial}{\partial\rho}\Big)(f-1)\Big(-\frac{\partial}{\partial x_0}+\frac{\partial}{\partial\rho}\Big)u=0,
\end{multline}
where  
\begin{equation}																				\label{eq:2.3}
f=1-\frac{2 m}{\rho}+\frac{e^2}{\rho^2}.
\end{equation}
We  have
$$
f=\frac{(\rho-r_+)(\rho-r_-)}{\rho^2},
$$
\begin{equation}																				\label{eq:2.4}
r_+= m+\sqrt{ m^2-e^2},\ \ r_-= m-\sqrt{ m^2-e^2},
\end{equation}
When  $e^2< m^2,\ r_+$  and $r_-$  are  distinct  real numbers  and  $\rho=r_+$  and  $\rho=r_-$  are  event  horizons.
The  RN  black  hole  is called  not-extremal in this case.  If  $e^2= m^2$  then
  \begin{equation}																				\label{eq:2.5}
  r_+=r_-=m
  \end{equation}
  is  the  only  event  horizon  and RN  hole  is  called  extremal.

We shall consider the Hawking radiation  in the exterior 
of the black hole  $\rho=r_+$  separately for  not-extremal  and extremal black holes.

First we shall briefly describe  the quantum  field theory for curved spacetime (cf. [8])
when the background  metric  is not-extremal or extremal  RN metric.  Note that the symbol  of  (\ref{eq:2.2})  is
\begin{equation}																				\label{eq:2.6}
(2-f)\eta_0^2+2(f-1)\eta_0\eta_\rho  -  f\eta_{\rho}^2-\frac{1}{\rho^2}m'^2.
\end{equation}
We shall introduce,  as in [3],   an ``orthonormal  basis"  of solutions of  the wave  equation  (\ref{eq:2.2})
and we will use the same notations  as in [3].

Let  $f_k^\pm(x_0,\rho,\varphi)$  be the solutions  of (\ref{eq:2.2})  with the initial  conditions
\begin{align}																				\label{eq:2.7}
&f_k^\pm(x_0,\rho,\varphi)\big|_{x_0=0}=\gamma_k e^{ik\cdot x}
\\
																						\label{eq:2.8}
&\frac{\partial f_k^\pm(x_0,\rho,\varphi)}{\partial x_0}\big|_{x_0=0}=i\lambda_0^\mp(k)\gamma_ke^{ik\cdot x}
\end{align}																						
where  $k=(\eta_\rho,m'),\ x=(\rho,\varphi),\ k\cdot x=\rho\eta_\rho+m'\varphi,$  and (cf. [3])
\begin{equation} 																			\label{eq:2.9}
\gamma_k=\frac{1}{\sqrt \rho\big(\eta_\rho^2+a^2\big)^{\frac{1}{4}}}\ \frac{1}{\sqrt 2\, 2\pi},
\end{equation}
\begin{equation}																			\label{eq:2.10}
\lambda_0^\pm(k)=\frac{-(f-1)\eta_\rho\pm \sqrt {\eta_\rho^2+a^2}}{2-f},
\end{equation}
where $a$  is arbitrary.

Note that  
\begin{equation}																	 		\label{eq:2.11}
f_k^+(x_0,\rho,\varphi)=\overline{f_{-k}^-(x_0,\rho,\varphi)},
\end{equation}
since  $-\lambda_0^+(-k)=\lambda_0^-(k)$.

Introduce  the Klein-Gordon (KG) inner product of any  two solutions  $u,v$  of (\ref{eq:2.2}):
\begin{equation}																			\label{eq:2.12}
<u,v>=i\int\limits_{x_0=t}g^{\frac{1}{2}}\sum_{j=0}^2g^{0j}\Big(\bar u\frac{\partial v}{\partial x_j}-\frac{\partial\bar u}{\partial x_j}v\Big) dx_1dx_2.
\end{equation}
Note that (\ref{eq:2.12}) is independent  of  $t$  (cf.  [8]) and in polar coordinates  we have  (cf. (\ref{eq:2.2})  and (\ref{eq:2.6}))
\begin{equation}																				\label{eq:2.13}
<u,v>=i\int\limits_{x_0=t}(2-f)\Big(\bar u\frac{\partial v}{\partial x_0}-\frac{\partial \bar u}{\partial x_0}v\Big) 
+(f-1)\Big(\bar u\frac{\partial v}{\partial\rho}-\frac{\partial\bar u}{\partial\rho}v\Big)\rho d\rho d\varphi.
\end{equation}
It follows from (\ref{eq:2.10})  that
\begin{equation}																				\label{eq:2.14}
(2-f)\frac{\partial f_k^+}{\partial x_0}+(f-1)\frac{\partial f_k^+}{\partial\rho} =-i\sqrt{\eta_\rho^2+a^2}f_k^+
+(f-1)\frac{1}{2\rho}f_k^+
\end{equation}
Thus,  similarily to [3],  we  have
\begin{multline} 																				\label{eq:2.15}
<f_k^+,f_{k'}^+>=\delta(k-k'),\ \ <f_{-k}^-,f_{-k'}^->=-\delta(k-k'),\ \ <f_k^+,f_{-k'}^->=0\\ \mbox{for all}\ \ k=(\eta_\rho,m),k'=(\eta_\rho',m'),
\end{multline}
where
\begin{equation}																				\label{eq:2.16}
\delta(k-k')=\delta(\eta_\rho-\eta_\rho')\delta_{mm'}.
\end{equation}
Note  that  terms containing  $\frac{1}{2\rho}$  in  (\ref{eq:2.15})  are canceled  since   $\frac{1}{2\rho}$  is real.
It follows from  (\ref{eq:2.15}),  (\ref{eq:2.16})  that
$$
\{f_k^+(x_0,\rho,\varphi),f_{-k}^-(x_0,\rho,\varphi)\}
$$
form  an ``orthonormal" basis  of solutions of (\ref{eq:2.2}).

Having  the basis $\{f_k^+,f_{-k}^-\}$  we  can  expend  any  solution  $C(x_0,\rho,\varphi)$  of  (\ref{eq:2.2})   as
\begin{equation}																				\label{eq:2.17}
C(x_0,\rho,\varphi)=\sum_{m=-\infty}^\infty\int\limits_{-\infty}^\infty\big(C^+(k)f_k^+(x_0,\rho,\varphi)+C^-(k)f_{-k}^-\big)d\eta_\rho,
\end{equation}
where 
\begin{equation}																				\label{eq:2.18}
C^+(k)=<f_k^+,C>,\ \ C^-(k)=- <f_{-k}^-,C>.
\end{equation}
We shall call $C(x_0,\rho,\varphi)$  a wave packet.

Analogously,  when $\Phi$  is the  field operator  (cf. [8]),  we have
\begin{equation}																				\label{eq:2.19}
\Phi=\sum_{m=-\infty}^\infty\int\limits_{-\infty}^\infty\big(\alpha_k^+f_k^++\alpha_{-k}^-f_{-k}^-\big)d\eta_\rho,
\end{equation}
where
\begin{equation}																				\label{eq:2.20}
\alpha_k^+=<f_k^+,\Phi>,\ \ \alpha_{-k}^-=- <f_{-k}^-,\Phi>.
\end{equation}
Operators   $\alpha_k^+,\alpha_{-k}^-$   are called  the annihilation and the  creation operators, respectively.  They satisfy   commutation 
relations  (cf. [8]):
\begin{equation}																				\label{eq:2.21}
[\alpha_k^+,\alpha_{-k'}^-]=\delta(k-k')I,
\ \ 
[\alpha_k^+,\alpha_{k'}^+]=0,\ \ [\alpha_{-k}^-,\alpha_{-k'}^-]=0,\ \ 
\end{equation}
$I$  is the identity operator.

Let
\begin{equation}																\label{eq:2.22}
C^\pm=\sum_{m=-\infty}^\infty\int\limits_{-\infty}^\infty C^\pm(k)f_{\pm k}^\pm(x_0,x) d\eta_\rho,
\end{equation}

Thus 
\begin{equation}     																\label{eq:2.23}
C=C^+  +C^-.
\end{equation}

It follows from  (\ref{eq:2.17}),  (\ref{eq:2.19})   that 
\begin{equation}																\label{eq:2.24}
<C,\Phi>=\sum_{m=-\infty}^\infty\int\limits_{-\infty}^\infty\big( \overline {C^{+}}(k)\alpha_k^{+}
-\overline {C^{-}}(k)\alpha_{-k}^{-}
\big) d\eta_\rho.
\end{equation}

 The vacuum  state $| 0\rangle$  is defined  (cf. [7])
by the conditions
\begin{equation}														\label{eq:2.25}
\alpha_k^{+}| 0\rangle  =0\ \ \mbox{for all\ }  k,
\end{equation}
i.e.  $|0\rangle$  annihilates all annihilation operators.

Let   $N(C)$  be  the number operator of particles    created  by  the wave  packet $C$ (cf.  [8])
\begin{equation}														\label{eq:2.26}
N(C)=<C,\Phi>^*<C,\Phi>,
\end{equation}
and let  $\langle 0| N(C) |0 \rangle$   be the average number of  particles.

As in [8]  one have the following theorem  (see Theorem 2.1  in [3]).
\begin{theorem}														\label{theo:2.1}
The average  number of particles  created  by the wave packet  $C$  is given by the formula
\begin{equation}														\label{eq:2.27}
\langle 0| N(C) | 0\rangle=\sum_{m=-\infty}^\infty\int\limits_{-\infty}^\infty| C^{-}(k)|^2 d\eta_\rho,
\end{equation}
where $C^{-}$  is the same  as in (\ref{eq:2.22}).
\end{theorem}

\section{The case of Hawking radiation  for non-extremal RN black hole in $|0\rangle$ vacuum state}
\init

First,  we find the solution  of the eikonal equation  
\begin{equation}																			\label{eq:3.1}
\big(2-f(\rho)\big)\hat S_{x_0}^2+2(f(\rho)-1\big)\hat S_{x_0} \hat S_\rho -f(\rho)\hat S_\rho^2-\frac{1}{\rho^2}\hat S_\varphi^2=0
\end{equation}
that tends to $\infty$    when  $\rho\rw r_+$. 

 We assume  $\hat S_{x_0}=-\eta_0,\ \hat S_\varphi=m, \eta_0>0$.  Then
\begin{multline}																				\label{eq:3.2}
\hat S_\rho^\pm=\frac{
\big(f(\rho)-1\big)\eta_0\pm\sqrt{
(f-1)^2\eta_0^2+(2-f)f\eta_0^2-\frac{1}{\rho^2}m^2f}
}
{-f}
\\
=\frac{\big(f(\rho)-1\big)\eta_0\pm \sqrt{\eta_0^2-\frac{f}{\rho^2}m^2}}{-f}.
\end{multline}
Therefore
\begin{equation}																			\label{eq:3.3}
\hat S_\rho^-=\frac{(f(\rho)-1)\eta_0-\sqrt{\eta_0^2-f\frac{m^2}{\rho^2}}}{-f}
=\frac{-2\eta_0}{
{-\frac{(\rho-r_+)(r_+-r_-)}{r_+^2}}
} +O(\rho-r_+),
\end{equation}
so
\begin{equation}																			\label{eq:3.4}
\hat  S_\rho^-=\frac{2\eta_0 r_+^2}{(r_+-r_-)(\rho-r_+)}+O(\rho-r_+).
\end{equation}
Thus,
\begin{equation}																			\label{eq:3.5}
\hat  S^-=-\eta_0x_0+\frac{2\eta_0r_+^2}{r_+-r_-}\ln|\rho-r_+|+m\varphi+O(\rho-r_+)\ \ \mbox{for}\ \ \rho>r_+.
\end{equation}
As in [3]  we define  the wave packet  $ \hat  C(x_0,\rho,\varphi)$  as  the exact  solution  of the wave  equation  (\ref{eq:2.2})  having the
 following initial  data  at $x_0=0$:
\begin{equation} 																			\label{eq:3.6}
\hat C\big|_{x_0=0}=\theta(\rho-r_+)\frac{1}{\sqrt \rho}(\rho-r_+)^\e e^{-a(\rho-r_+)}\exp{i\hat\xi_0\ln|\rho-r_+|+im\varphi},
\end{equation}
\begin{equation}																			\label{eq:3.7}
\frac{\partial \hat C}{\partial x_0}\Big|_{x_0=0}=i\hat\beta \hat C\big|_{x_0=0},
\end{equation}
where  $a>0,  \e>0$,
\begin{equation}	 																		\label{eq:3.8}
\hat\xi_0=\frac{2\eta_0r_+^2}{r_+-r_-},
\end{equation}
\begin{equation}																			\label{eq:3.9}
 \hat\beta=\frac{-f\frac{\hat\xi_0}{\rho-r_+}}{2-f}.
\end{equation}
For  the convenience we take  $a>0$  in  (\ref{eq:3.6}) equal  to $a>0$  in (\ref{eq:2.9}).
Note that 
\begin{multline}																				\label{eq:3.10}
\hat\beta=\frac{-f  \frac{\hat\xi_0}{\rho-r_+}}{2-f}=
\frac{\frac{-(\rho-r_+)(\rho-r_-)}{r_+^2}\cdot \frac{2\eta_0 r_+^2}{(r_+-r_-)(\rho-r_+)}}{2}+O(\rho-r_+)
\\
=-\eta_0+O(\rho-r_+).
\end{multline}

  Calculating  the KG norm of $\hat C$,  we get ,  as in    [3],
\begin{equation}                          																\label{eq:3.11}
<\hat C,\hat C>=\frac{4\pi\hat\xi_0\Gamma(2\e)}
{(2a)^\e}.
\end{equation}
Note that  the terms containing  the derivative of $\frac{1}{\sqrt \rho}(\rho-r_+)^\e e^{-a(\rho-r_+)}$ are canceled since they are the derivatives
of a real-valued function.

As in (\ref{eq:3.4})  in [5],   we have
\begin{equation}								                 									\label{eq:3.12}
\hat C^{-}(k)=- < f_{-k}^{-},\ \ \hat C>=\hat C_1^{-}(k)+\hat C_2^{-}(k),
\end{equation}
where
\begin{equation} 																			\label{eq:3.13}
\hat C_1^{-}(k)=-i\int\limits_0^\infty \int\limits_0^{2\pi}
\Big( (2-f)\frac{\partial  f_k^{+}}{\partial x_0}
+(f-1)\frac{\partial  f_k^{+}}{\partial\rho}\Big)
\,\hat  C\rho  d\rho  d\varphi
\end{equation}
and
\begin{equation} 																			\label{eq:3.14}
\hat C_2^{-}(k)=
i\int\limits_0^\infty \int\limits_0^{2\pi}
 f_k^{+}\Big( (2-f)\frac{\partial \hat C}{\partial x_0}
+(f-1)\frac{\partial \hat C}{\partial\rho}\Big)
\,  \rho d\rho  d\varphi
\end{equation}
Note that for $x_0=0$  
\begin{align}																				\label{eq:3.15}
&\Big( (2-f)\frac{\partial \hat C}{\partial x_0}
+(f-1)\frac{\partial \hat C}{\partial\rho}\Big)
\\
\nonumber
&
=\Big[-\frac{i\hat\xi_0}{\rho-r_+}+(f-1)\big(\frac{\e}{\rho-r_+}-a-\frac{1}{2\rho}\big)\Big]\,\hat C\big|_{x_0=0},
\end{align}
where we used that
\begin{equation}																			\label{eq:3.16}
(2-f)i\hat\beta +(f-1)\frac{i\hat\xi_0}{\rho-r_+}=-\frac{i\hat\xi_0}{\rho-r_+}.
\end{equation}

Using  
 (\ref{eq:3.13}),  (\ref{eq:3.14})  we  have  (cf.  (3.5),  (3.6)  in  [5]):
\begin{multline}																\label{eq:3.17}
\hat C_2^-(k)
=\int\limits_0^\infty\int\limits_0^{2\pi}\frac{e^{i\rho\eta_\rho+im'\varphi}\theta(\rho-r_+)}
{\sqrt\rho (\eta_\rho^2+a^2)^{\frac{1}{4}}\sqrt 2 \ 2\pi}
\ \frac{(\rho-r_+)^\e}{\sqrt \rho}\ 
\Big(\frac{\hat\xi_0}{\rho-r_+}-(f-1)\Big(\frac{-i\e}{\rho-r_+}+ia\Big)\Big)
\\
\cdot e^{-a(\rho-r_+)+i\hat\xi_0\ln(\rho-r_+)}\
  e^{im\varphi} \rho\, d\rho d\varphi,
\end{multline}
\begin{multline}																\label{eq:3.18}
\hat C_1^-(k)
=-\int\limits_0^\infty\int\limits_0^{2\pi}\frac{e^{i\rho\eta_\rho+im'\varphi}(\eta_\rho^2+a^2)^{\frac{1}{4}}
}
{\sqrt\rho \sqrt 2 \ 2\pi}
\ \frac{\theta(\rho-r_+)(\rho-r_+)^\e}{\sqrt \rho}\ 
\\
\cdot e^{-a(\rho-r_+)+i\hat \xi_0\ln(\rho-r_+)+im\varphi}\
 \rho d\rho d\varphi,\ \ \ k=(\eta_\rho,m').
\end{multline}
Note that the terms  in $\hat C_1^-+\hat C_2^-$  containing $\frac{1}{2\rho}$  are cancelled  (cf.  (\ref{eq:2.15})).

Integrating  in  $\varphi$  and using the  formula  (cf.  [5])
\begin{equation}															\label{eq:3.19}
\int\limits_0^\infty  e^{it\eta_\rho}t^\lambda e^{-at}dt  =\frac{e^{i\frac{\pi}{2}(\lambda+1)}\Gamma(\lambda+1)}
{(\eta_\rho+ia)^{\lambda+1}},
\end{equation}
we get
\begin{multline}															\label{eq:3.20}
\hat C_2^-(k)=\frac{\delta_{m',-m}}{\sqrt 2}e^{ir_+\,\eta_\rho}
\frac{ e^{i\frac{\pi}{2}(i\hat\xi_0+\e)}\Gamma(i\hat\xi_0+\e)(\hat\xi_0 -i\e)}
{(\eta_\rho^2+a^2)^{\frac{1}{4}}(\eta_\rho+ia)^{i\hat\xi_0+\e}} 
\\
+\frac{\delta_{m',-m}}{\sqrt 2}e^{ir_+\, \eta_\rho}
\frac{ia e^{i\frac{\pi}{2}(i\hat\xi_0+\e+1)}\Gamma(i\hat\xi_0+\e+1)}
{(\eta_\rho^2+a^2)^{\frac{1}{4}}(\eta_\rho+ia)^{i\hat\xi_0+\e+1}} 
 +\delta_{m',-m}O\Big(\frac{1}{|\eta_\rho+ia|^{1+\e}}\Big),
\end{multline}
\begin{equation}															\label{eq:3.21}
C_1^-(k)=-\frac{\delta_{m',-m}}{\sqrt 2}e^{ir_+\eta_\rho}
\frac{e^{i\frac{\pi}{2}(i\hat\xi_0+\e+1)}\Gamma(i\hat\xi_0+\e+1)(\eta_\rho^2+a^2)^{\frac{1}{4}}
}
{
(\eta_\rho+ia)^{i\hat\xi_0+\e+1}},
\end{equation}
where $\delta_{m_1,m_2}=1$  when  $m_1=m_2$  and  $\delta_{m_1,m_2}=0$  when   $m_1\neq m_2$.                                                                                                                                                                                                                                                                                                                                                                                                                                                                                                                                                                                                                                                                                                                                                                                                                                                                                                                                                                                                                                                                                                                                                                                                                                                                                                                                                                                                                                                                                                                                                                                                                                                                                                                                                                                                                                                                                                                                                                                                                                                                                                                                                                                                                                                                                                                                                                                                                                                                                                                                                                                                                                                                                                                                                                                                                                                                                                                                                                                                                                                                                                                                                                                                                                                                                                                                                                                                                                                                                                                                                                                        
Note   that   $\big| e^{i\frac{\pi}{2}(i\hat\xi_0+\e)}\big|
=e^{-\frac{\pi}{2}\hat\xi_0},\ \Gamma(i\hat\xi_0+\e+1)=(i\hat\xi_0+\e)e^{-\frac{\pi}{2}\hat\xi_0}\Gamma_1(i\hat\xi_0+\e),\linebreak \Gamma_1$  is bounded,
$\Gamma_1=i\int\limits_0^\infty e^{(i\hat\xi_0+\e-1)\ln y+i(\e-1)\frac{\pi}{2}-iy}dy. $    
We have
\begin{multline}															\label{eq:3.22}
|\hat C^-|^2=|\hat C_1^- +\hat  C_2^-|^2
\\
=
\frac{\delta_{m',-m}}{2}
\Big|\frac{\hat\xi_0-i\e}{(\eta_\rho^2+a^2)^{\frac{1}{4}}}
+\frac{ia(-\xi_0+i\e)}{(\eta_\rho^2+a^2)^{\frac{1}{4}}(\eta_\rho +ia)}
-\frac{i(i\hat \xi_0+\e)(\eta_\rho^2+a^2)^{\frac{1}{4}}}{\eta_\rho+ia}\Big|^2
\\
\cdot e^{-2\pi\hat \xi_0}
|\Gamma_1(i\hat\xi_0+\e)|^2e^{2\hat\xi_0\arg(\eta_\rho+ia)}
(\eta_\rho^2+a^2)^{-\e}
+\delta_{m',-m}O\big(|\eta_\rho+ia|^{-2-2\e}\big)
\\
=\frac{ \delta_{m',-m}}{2}|\hat\xi_0-i\e|^2 \Big|\frac{\eta_\rho}{(\eta_\rho^2+a^2)^{\frac{1}{2}}}  +(\eta_\rho^2+a^2)^{\frac{1}{4}}\Big|^2
\\
\cdot e^{-2\pi\hat \xi_0}
|\Gamma_1(i\hat\xi_0+\e)|^2e^{2\hat\xi_0\arg(\eta_\rho+ia)}|\eta_\rho+ia|^{-2\e-2}
\\
+\delta_{m',-m}O\big(|\eta_\rho+ia|^{-2-2\e}\big).
\end{multline}
Therefore integrating  in $\eta_\rho$,  summing in $m'$  
and
 changing  $\eta_\rho=a\eta_\rho'$   we  get  
\begin{equation}                       											\label{eq:3.23}
\langle 0|N(\hat C)|0\rangle =a^{-2\e}\int\limits_{-\infty}^\infty \hat C_3(\eta_\rho')d\eta_\rho' +O\big(a^{-2\e-1}\big),
\end{equation}
where
\begin{multline}														\label{eq:3.24}
\hat C_3(\eta_\rho)=\frac{1}{2}e^{-2\pi\hat\xi_0}
|\Gamma_1(i\hat\xi_0+\e)|^2\, |\hat\xi_0+i\e|^2
\ \Big|\frac{\eta_\rho}{(\eta_\rho^2+1)^{\frac{1}{4}}}+(\eta_\rho^2+1)^{\frac{1}{4}}\Big|^2
\\
\cdot e^{2\hat\xi_0 
\arg(\eta_\rho+i)}
(\eta_\rho^2+1)^{-\e-1}.
\end{multline}
Let   $\hat C_n=\frac{\hat C}{<\hat C,\hat C>^{\frac{1}{2}}}$.    Thus   $<\hat C_n,\hat C_n>=1$.
Therefore, taking into account    (\ref{eq:3.11}),  we get:
\begin{equation}													\label{eq:3.25}
\lim_{a\rw \infty} \langle 0| N(\hat C_n)| 0\rangle=
\frac{2^\e }{4\pi \Gamma(2\e)}\int\limits_{-\infty}^\infty\frac{1}{\hat\xi_0}\hat C_3(\eta_\rho)d\eta_\rho.
\end{equation}
Note that  $\hat C_3(\eta_\rho)$  in  (\ref{eq:3.25})  is identical  to the  $C_3(\eta_\rho)$  in (\ref{eq:3.13}) in  [5] 
 after we replace  $\xi_0|A|$  by $\hat\xi_0$.

Let
$$
I_+=\frac{2^\e}{4\pi\Gamma(\e)}\int\limits_0^\infty\frac{1}{\hat\xi_0}\hat C_3(\eta_\rho)d\eta_\rho,\ \
I_-=\frac{2^\e}{4\pi\Gamma(\e)}\int\limits_{-\infty}^0\frac{1}{\hat\xi_0}\hat C_3(\eta_\rho)d\eta_\rho,
$$
i.e.
\begin{equation}							 			 			\label{eq:3.26}
\lim_{a\rw\infty}\langle 0|N(\hat C_n)|0\rangle=I_++I_-.
\end{equation}
As it  is shown  in the end  of  \S 3  in [5],   $I_+=O(e^{-\pi\hat\xi_0})$   and   $I_-=O(\hat\xi_0^{1-\delta(1+2\e)}),\linebreak 0<\delta<1$.

The integral  $I_+$  gives  the contribution to the Hawking radiation  and the contribution  of $I_-$  is unrelated to the Hawking radiation.

\section{The  Hawing radiation for  not-extremal RN black hole in  the Unruh type vacuum}
\init

Consider the equation  (\ref{eq:2.2})  
with $r_+ > r_-$,  i.e.  the  case  of not-extremal  RN equation.

When using  the vacuum state $| 0\rangle$   we get  that  the average  number  of created particles  consists  of two parts:
the part  related to the Hawking radiation  and the part that is not related.  In this section  we  introduce  another vacuum state  that is  called
the Unruh type vacuum state.  The average number of particles  in this vacuum state  consists only  of particles  related  to the Hawking radiation.

To introduce  the Unruh  type vacuum state,   we will split,  as in [3]:
\begin{align}																				\label{eq:4.1}
&f_k^{++}=f_k^+ \theta(\eta_\rho),\ \ \ f_k^{+-}=f_k^+\big(1-\theta(\eta_\rho)\big),
\\																						\label{eq:4.2}
&f_{-k}^{-+}=f_{-k}^- \theta(\eta_\rho),\ \ \ f_{-k}^{--}=f_{-k}^-\big(1-\theta(\eta_\rho)\big),
\end{align}  

where  $\theta(\eta_\rho)=1$  when  $\eta_\rho>0,\ \theta(\eta_\rho)=0$  when  $\eta_\rho<0$. 
Analogously,  we define  $\alpha_k^{++},\alpha_k^{+-}, \alpha_{-k}^{-+},\alpha_{-k}^{--}$.   Therefore
\begin{equation}																				\label{eq:4.3}
\Phi=\sum_{m=-\infty}^\infty\int\limits_{-\infty}^\infty\big(
\alpha_k^{++}f_k^{++}+\alpha_k^{+-}f_k^{+-}+\alpha_{-k}^{-+}f_{-k}^{-+}+\alpha_{-k}^{--}f_{-k}^{--}\big)d\eta_\rho,
\end{equation}
and
\begin{equation}																				\label{eq:4.4}
C=
\sum_{m=-\infty}^\infty\int\limits_{-\infty}^\infty
C^{++}(k)f_k^{++}+C^{+-}(k)f_k^{+-}+C^{-+}(k)f_{-k}^{-+}+C^{--}(k)f_{-k}^{--}\big)d\eta_\rho,
\end{equation}
where  (cf. (\ref{eq:4.1}))
\begin{align}																				\label{eq:4.5}
&C^{++}(k)=C^+ \theta(\eta_\rho),\ \ C^{+-}(k)=C^+(1-\theta(\eta_\rho),
\\
\nonumber
&C^{-+}(k)=C^-\theta(\eta_\rho),\ \ C^{--}=C^-(1-\theta(\eta_\rho)).
\end{align}
As in  [3]   (cf. [10]),   we define  the Unruh type vacuum  state  $|\Psi\rangle$  by the  conditions
\begin{equation}																		\label{eq:4.6}             
\alpha_k^{++}|\Psi\rangle=0,\ \ \ \alpha_{-k}^{--}|\Psi\rangle=0,\ \ \forall k.
\end{equation}
It follows  from  (\ref{eq:4.3}),  (\ref{eq:4.4})  that
\begin{equation}																				\label{eq:4.7}       
<C,\Phi>=
\sum_{m=-\infty}^\infty\int\limits_{-\infty}^\infty
\big(\overline{C^{++}(k)}\alpha_k^{++}+\overline{C^{+-}(k)}\alpha_k^{+-}-
\overline{C^{-+}(k)}\alpha_{-k}^{-+}-\overline{C^{--}(k)}\alpha_{-k}^{--}\big)d\eta_\rho,
\end{equation}
The  operator  of the  number  of particles created  by the wave  packet  $C$  is  equal  to  (cf.  [8])
\begin{equation}																				\label{eq:4.8}       
N(C)=<C,\Phi>^*<C,\Phi>.
\end{equation}
The average  number  of created  particles  is 
$$
\langle \Psi|N(C)|\Psi\rangle.
$$
Exactly  as in  [3]  we  get  (see Theorem  2.1  in  [3])
\begin{equation}																				\label{eq:4.9}       
\langle\Psi | N(C) |\Psi\rangle=\sum_{m=-\infty}^\infty\int\limits_{-\infty}^\infty \Big( -|C^{+-}(k)|^2+|C^{-+}(k)|^2\Big)d\eta_\rho.
\end{equation}
  Let  $\hat C(x_0,\rho,\varphi)$  be  the same  wave  packet  as in \S 3  (see  (\ref{eq:3.6}),   (\ref{eq:3.7}),   (\ref{eq:3.8}),   (\ref{eq:3.9})).
  
  To compute the average number  of particles  created by  the wave  packet $\hat C$  we need  to compute  $\hat C^{+-}(k)$  
  and  $\hat C^{-+}(k)$.
$\hat C^{-+}(k)$  was computed  in \S 3.  Thus it remains  to compute  $\hat C^{+-}(k)$.

We have 
\begin{equation}																				\label{eq:4.10}
\hat C^{+-}(k)=\hat C_1^{+-}(k)  +\hat C_2^{+-}(k),
\end{equation}
where  $\hat C^{+-}(k)=<f_k^{+-},\hat C>$.

Analogously  to (\ref{eq:3.13}),  (\ref{eq:3.14}),  we have
\begin{equation}																				\label{eq:4.11}
\hat C_1^{+-}(k)=i\int\limits_0^\infty\int\limits_0^{2\pi}\Big((2-f)
\frac{\overline{\partial f_k^{+-}}}
 {\partial x_0}+ (f-1)
\frac{\overline{\partial f_k^{+-}}}{\partial \rho}\Big)\hat C\rho d\rho d\rho d\varphi,
\end{equation}
\begin{equation}																				\label{eq:4.12}
\hat C_2^{+-}(k)=-i\int\limits_0^\infty\int\limits_0^{2\pi}\overline{ f_k^{+-}}\Big((2-f)
\frac{\partial \hat C}
 {\partial x_0}+ (f-1)
\frac{\partial \hat C}{\partial \rho}\Big)\rho d\rho d\rho d\varphi.
\end{equation}
Therefore  $\hat C_1^{+-}$  and  $\hat C_2^{+-}$  have  the form:
\begin{multline} 																				\label{eq:4.13}
\hat C_1^{+-}(k)=-\int\limits_0^\infty\int\limits_0^{2\pi}
\frac{(\eta_\rho^2+a^2)^{\frac{1}{2}}}{2\pi\sqrt 2\sqrt\rho}e^{-i\rho\eta_\rho-im'\varphi}
\\
\cdot\theta(\rho-r_+)\frac{(\rho-r_+)^\e}{\sqrt \rho}e^{-a(\rho-r_+)+i\hat\xi_0\ln(\rho-r_+)+im\varphi}  \rho d\rho d\varphi,
\end{multline}
\begin{multline}																					\label{eq:4.14}
\hat C_2^{+-}(k)=-\int\limits_0^\infty\int\limits_0^{2\pi}
\frac{e^{-i\rho\eta_\rho-im'\varphi}}{2\pi\sqrt 2\sqrt \rho (\eta_\rho^2+a^2)^{\frac{1}{2}}}
\theta(\rho-r_+)(\rho-r_+)^\e e^{-a(\rho-r_+)}
\\
\cdot e^{i\hat\xi_0\ln(\rho-r_+)+im\varphi}
\Big(\frac{\hat\xi_0}{\rho-r_+}-(f-1)\Big(\frac{i\e}{\rho-r_+}-ia\Big)\Big)\rho d\rho d\varphi.
\end{multline}

Integrating  in $\varphi$  and using  the formula  (cf. [3]))
\begin{equation}																				\label{eq:4.15}
\int\limits_{0}^\infty e^{-i\eta_\rho t}t^\lambda e^{-at}dt=\frac{\Gamma(\lambda+1)e^{-i\frac{\pi}{2}(\lambda+1)}}{(\eta_\rho-ia)^{\lambda+1}},
\end{equation}
we get
\begin{equation}																				\label{eq:4.16}
\hat C_1^{+-}(k)=-\frac{\delta_{m,-m'}}{\sqrt 2}
\frac
{
(\eta_\rho^2+a^2)^{\frac{1}{4}}e^{-ir_+\, \eta_\rho}\Gamma(i\hat\xi_0+\e+1)
}
{
(\eta_\rho -ia)^{i\hat\xi_0+\e+1}
}
e^{-\frac{i\pi}{2}(i\hat\xi_0+\e+1)},
\end{equation}
\begin{equation}																				\label{eq:4.17}
\hat C_2^{+-}(k)=-\frac{\delta_{m,-m'}}{\sqrt 2}
e^{-ir_+\eta_\rho}
\frac
{
\Gamma(i\xi_0+\e)e^{-\frac{i\pi}{2}(i\hat\xi_0+\e)}
}
{
(\eta_\rho^2+a^2)^{\frac{1}{4}}
(\eta_\rho -ia)^{i\hat\xi_0+\e}
}
\Big(\hat\xi_0-i\e+\frac{ia e^{-i\frac{\pi}{2}}(i\hat\xi_0+\e)}{\eta_\rho-ia}\Big)
\end{equation}

Note that  the terms  $-\frac{1}{2\rho}\overline{\hat f_k^{+-}}\hat C$  in  (\ref{eq:2.12})  are canceled.

Therefore,  for  $\eta_\rho<0$  we  have
\begin{multline}																						\label{eq:4.18}
|\hat C_1^{+-}+\hat C_2^{+-}|^2=\frac{\delta_{m,-m'}}{2}\frac{|\Gamma_1(i\hat \xi_0+\e)|^2
e^{   \hat\xi_0\big(-2\pi+2\sin^{-1}\frac{a}{\sqrt{\eta_\rho^2+a^2}}\big)     }
}
{|\eta_\rho-ia|^{2\e}}
\\
\cdot 
\Big|\frac
{    (\hat\xi_0-i\e)(\eta_\rho^2+a^2)^{\frac{1}{4}}     }{\eta_\rho-ia }  
+\frac{    \hat\xi_0 -i\e    }{     (\eta_\rho^2+a^2)^{\frac{1}{4}}       }
\Big(1+
\frac{ia}{\eta_\rho-ia}
\Big)\Big|^2+
O\Big(\frac{1}{     (\eta_\rho-ia)^{2\e+2}     }
\Big)
\\
=\frac{\delta_{m,-m'}}{2}
\frac
{|\Gamma_1(i\hat \xi_0+\e)|^2
e^{   -2\pi\hat\xi_0+2\hat\xi_0\sin^{-1}\frac{a}{     \sqrt{\eta_\rho^2+a^2}    }  }
   }
{|\eta_\rho-ia|^{2\e+2}} |\hat\xi_0 -i\e|^2
\\
\cdot 
\Big|
(\eta_\rho^2+a^2)^{\frac{1}{4}}   
+\frac{\eta_\rho}
{(\eta_\rho^2+a^2)^{\frac{1}{2}}}
\Big|^2
+
O\Big(\frac{1}{     (\eta_\rho-ia)^{2\e+2}     }
\Big)
\end{multline}
Note that
\begin{equation}																					\label{eq:4.19}
\Big|
(\eta_\rho^2+a^2)^{\frac{1}{4}}   
+\frac{\eta_\rho}
{(\eta_\rho^2+a^2)^{\frac{1}{2}}}
\Big|^2
-
\Big|
(\eta_\rho^2+a^2)^{\frac{1}{4}}   
+\frac{-|\eta_\rho|}
{(\eta_\rho^2+a^2)^{\frac{1}{2}}}
\Big|^2
=4\eta_\rho
\end{equation}
 when  $\eta_\rho>0$.  Therefore
\begin{multline}																				\label{eq:4.20}
\int\limits_0^\infty|C^{-+}|^2d\eta_\rho
- \int\limits_{-\infty}^0|C^{+-}|^2d\eta_\rho
\\
=
2\delta_{m,-m'}|\hat\xi_0-i\e|^2
e^{-2\pi\hat\xi_0}|\Gamma_1|^2
\int\limits_0^\infty\frac{\eta_\rho}{(\eta_\rho^2+a^2)^{\e+1}}
e^{ 2\hat\xi_0\sin^{-1}\frac{a}{     \sqrt{\eta_\rho^2+a^2}    }  }
d\eta_\rho
+ O(a^{-2\e-1}).
\end{multline}

  Finally,  normalize  $\hat C$,  i.e.   replace  $\hat C$  by
\begin{equation}																	\label{eq:4.21}                   
\hat C_n=\frac{\hat C}{<\hat C,\hat C>^{\frac{1}{2}}},
\end{equation}
 make change  of variables $\eta_\rho=a\eta_\rho'$  and  use  (\ref{eq:3.11}).
Then we will get,  as in Theorem 3.1 in [3],  that
\begin{multline}																		\label{eq:4.22}                   
\lim_{a\rw\infty}  \langle \Psi |N(\hat C_n)|\Psi\rangle
=\frac{2^\e\, e^{-2\pi\hat\xi_0}|\Gamma_1|^2}{2\pi\Gamma(\e)}
\\
\cdot 
\frac{(\hat\xi_0^2+\e^2)}{\hat\xi_0}\int\limits_{-\infty}^0\frac{|\eta_\rho|}{(\eta_\rho^2+1)^{\e+1}}
e^{2\hat\xi_0\sin^{-1}\frac{1}{\sqrt{\eta_\rho^2+1}}}d\eta_\rho.
\end{multline}
Therefore  we proved  that the Hawking  radiation  for  non-extremal  RN  
black hole  has the same  form  as  the Hawking  radiation 
 for the  rotating  acoustic  black  hole  with  
$\hat\xi_0=\frac{2\eta_0r_+^2}{r_+-r_-}$  replacing  $\xi_0|A|=\big(\eta_0-\frac{Bm}{|A|^2}\big)|A|$.

The proof given here is slightly  different  from the proof in [3]  since we wanted  to use the fact that 
$\int\limits_0^\infty|\hat C^{-+}|^2d\eta_\rho$  is already  computed  in \S 3.

\section{The   Hawking  radiation  for  extremal  RN  black hole}
\init

In the case  of two space dimensions  one have  the wave  equation  (\ref{eq:3.3})  when  $e^2= m^2$.  Thus
\begin{equation}																		\label{eq:5.1}
r_+=r_-= m.
\end{equation}
In particular,
\begin{equation}																		\label{eq:5.2}
f=1-\frac{2 m}{\rho}+\frac{e^2}{\rho^2}=\Big(1-\frac{ m}{\rho}\Big)^2.
\end{equation}
We shall  treat  the  extremal  RN  case following  the  prescription of \S 3.

We start  with the  eikonal  that tends  to $\infty$  when  $\rho \rw  m $ 
(cf.  (\ref{eq:3.3})).  We have
\begin{multline}																			\label{eq:5.3}
\tilde S_\rho^-=\frac{
\big( f(\rho)-1\big)\eta_0
-\sqrt{\eta_0^2-\frac{fm'^2}{\rho^2}}
}{-f}
\\
=\frac{-2\eta_0}{-\frac{(\rho- m^2)}{\rho^2} } +O(\rho- m)=\frac{2\eta_0 m^2}{(\rho- m)^2}+O(\rho- m),
\end{multline}
where
$\eta_0>0,\ m'\in \Z$.
Therefore  the eikonal
$\tilde S^-$  has the form
\begin{equation}																		\label{eq:5.4}
\tilde S^-=-\eta_0x_0-\frac{2\eta_0  m^2}{\rho- m}+O(\rho- m)+m'\varphi.
\end{equation}
As in \S3  we use  the approximation of  $\tilde S^-$  to  construct  the wave  packet  $\tilde C(x_0,\rho,\varphi)$  as the exact solution  of (\ref{eq:2.2})
with the following initial data
\begin{equation}																		\label{eq:5.5}
\tilde C\big|_{x_0=0}=\theta(\rho- m)\frac{1}{\sqrt\rho}(\rho- m)^\e
 e^{-a(\rho-m)}e^{-\frac{i\tilde\xi_0}{\rho- m}+im'\varphi},
\end{equation}
\begin{equation}																		\label{eq:5.6}
\frac{\partial\tilde C}{\partial x_0}\Big|_{x_0=0}=i\tilde\beta\tilde C\big|_{x_0=0},
\end{equation}
where
\begin{equation}																		\label{eq:5.7}
\tilde \xi_0=2\eta_0  m^2,
\end{equation}
\begin{equation}																		\label{eq:5.8}
\tilde\beta=\frac{-f\frac{\tilde\xi_0}{(\rho- m)^2}}{2-f}.
\end{equation}
Note  that
\begin{equation}																		\label{eq:5.9}
(2-f)\tilde\beta+(f-1)\frac{\tilde\xi_0}{(\rho- m)^2}=\frac{-\tilde\xi_0}{(\rho- m)^2}.
\end{equation}
Note also that
\begin{multline}																		\label{eq:5.10}
(2-f)\frac{\partial\tilde C}{\partial x_0}\Big|_{x_0=0}+(f-1)\frac{\partial \tilde C}{\partial\rho}
\\
=\Big[(2-f)i\tilde \beta +(f-1)\Big(\frac{i\tilde\xi_0}{(\rho- m)^2}+\frac{\e}{\rho- m}-a-\frac{1}{2\rho}\Big)\Big]
\tilde C\big|_{x_0=0}.
\end{multline}
Using  (\ref{eq:3.9}),  (\ref{eq:3.10})  we shall  
compute,  as in [3],   the KG   norm  of  $\tilde C$:
\begin{align}																				\label{eq:5.11}
<\tilde C,\tilde C>=
&\int\limits_0^{2\pi}\int\limits_m^\infty (\rho- m)^{2\e} e^{-2a(\rho- m)}\frac{2\tilde\xi_0}{(\rho-  m)^2}d\rho d\varphi
\\
\nonumber
&=4\pi\tilde\xi_0\int\limits_0^\infty t^{2\e -2}e^{-2at}dt=\frac{4\pi\tilde \xi_0}{(2a)^{2\e-1}}\Gamma(2\e-1),
\end{align}
where  we assume  that  $\e>\frac{1}{2}$.

Now we compute 
\begin{equation}																			\label{eq:5.12}
\langle 0|N(\tilde C)|0\rangle=
\sum_{m'=\infty}^\infty \int\limits_{-\infty}^\infty 
|C^-(k)|^2
d\eta_\rho.
\end{equation}
As in  (\ref{eq:3.10})  of  [3],  we have
\begin{equation}																			\label{eq:5.13}
\tilde C^{-}(k)=<\overline{f_k^{+}},\tilde C>=\tilde C_1^{-}(k)+\tilde C_2^{-}(k).
\end{equation}
Analogously to (\ref{eq:3.13})--(\ref{eq:3.16}),  we get
\begin{multline}																				\label{eq:5.14}
\tilde C_1^{-}(k)=
-\int\limits_0^\infty\int\limits_0^{2\pi}
\frac{\big(\eta_\rho^2+a^2\big)^{\frac{1}{4}}}{\sqrt 2\, 2\pi\sqrt\rho}
e^{i\rho\eta_\rho+im'\varphi}\theta(\rho- m)\frac{(\rho- m)^\e}{\sqrt\rho}e^{-a(\rho- m)}
\\
\cdot e^{-\frac{i\tilde\xi_0}{\rho-\hat m}+im''\varphi}\rho d\rho d\varphi,
\end{multline}
where   $m', m''$  are integers.
We used in (\ref{eq:5.14})  (cf.  (\ref{eq:2.15}))  that
\begin{equation}  																			\label{eq:5.15}
(2-f)\frac{\partial   f_k^{+}}{\partial x_0}+(f-1)\frac{\partial  f_k^{+}}{\partial\rho}=
-i\sqrt{\eta_\rho^2+a^2}
 f_{k}^{+}
-(f-1)\Big(-\frac{1}{2\rho}\Big)
 f_k^{+}
.
\end{equation}
Analogously,
\begin{multline}																			\label{eq:5.16}
\tilde C_2^{-}=\int\limits_0^\infty\int\limits_0^{2\pi}\frac{1}{\sqrt 2\, 2\pi\big(\eta_\rho^2+a^2\big)^{\frac{1}{4}}}
e^{i\rho\eta\rho+im'\varphi}\theta (\rho- m)(\rho- m)^\e
\\
\cdot 
e^{-a(\rho-m)}e^{-\frac{i\tilde\xi_0}{\rho- m}+im''\varphi}
\Big[\frac{\tilde\xi_0}{(\rho- m)^2}+i(f-1)\Big(\frac{\e}{\rho- m}-a-\frac{1}{2\rho}\Big)\Big] d\rho d\varphi.
\end{multline}
  In (\ref{eq:3.15})  and  (\ref{eq:3.16}) we will drop terms containing  $-\frac{1}{2\rho}$,  since  they will  cancel each other  
when we will  take the sum  $\tilde C_1^{-}$  and  $\tilde C_2^{-}$.
Integrate  in $\varphi$  in  (\ref{eq:5.14})  and  (\ref{eq:5.16})  and change  $\rho= m+t$. We  get

\begin{equation}													 					\label{eq:5.17}
\tilde C_1^{-}=\frac{-\delta_{m',-m''}\big(\eta_\rho^2+a^2\big)^{\frac{1}{4}}}{\sqrt 2}
e^{i m\eta_\rho}\, A_1(\eta_\rho,\tilde\xi_0,a),
\end{equation}
\begin{equation}													 					\label{eq:5.18}
\tilde C_2^{-}=\frac{\delta_{m',-m''}\big(\eta_\rho^2+a^2\big)^{-\frac{1}{4}}}{\sqrt 2}
e^{im\eta_\rho}\, A_2(\eta_\rho,\tilde\xi_0,a),
\end{equation}
where 
\begin{equation}																		\label{eq:5.19}
A_1(\eta_\rho,\tilde\xi_0,a)=\int\limits_0^\infty e^{it\eta_\rho}t^\e e^{-at-\frac{i\tilde\xi_0}{t}}dt,
\end{equation}
\begin{equation}																		\label{eq:5.20}
A_2(\eta_\rho,\tilde\xi_0,a)=\int\limits_0^\infty e^{it\eta_\rho}t^\e e^{-at-\frac{i\tilde\xi_0}{t}}
\Big[\frac{\tilde\xi_0}{t^2}+i(f-1)\Big(\frac{\e}{t}-a\Big)\Big]
dt,
\end{equation}
When $\eta_\rho>0$ make  in  (\ref{eq:5.19}),  (\ref{eq:5.20})  change  of variables
$$																		
(-i\eta_\rho+a)t =\tau.
$$
We get
\begin{equation}																		\label{eq:5.21}
A_1=(-i\eta_\rho+a)^{-\e-1}\int\limits_{\Gamma_{i\eta_\rho+a}} \tau^\e  e^{-\tau}e^{\frac{(-\tilde\xi_0\eta_\rho-ia\tilde\xi_0}{\tau}} d\tau,
\end{equation}
where
$\Gamma_{i\eta_\rho+a}$
is  the ray  $\{(-i\eta_\rho+a)t,t\geq 0\}$.

Using  the Cauchy  integral  theorem  we transform  (\ref{eq:5.21})  to the  integral  over  positive  semi axis:
\begin{equation}																		\label{eq:5.22}
A_1=(-i\eta_\rho+a)^{-\e-1}\int\limits_0^\infty \tau^\e  e^{-\tau}e^{\frac{-\tilde\xi_0\eta_\rho-ia\tilde\xi_0}{\tau}} d\tau,
\end{equation}
Note  that  $-\tilde\xi_0\eta_\rho<0$  since  $\eta_\rho>0$.
Changing  coordinates  as  in  (\ref{eq:5.21})  and  using  the Cauchy  theorem,  we get from  (\ref{eq:5.20})
\begin{multline}																		\label{eq:5.23}
A_2
=\tilde\xi_0(-i\eta_\rho+a)^{-\e+1}\int\limits_0^\infty \tau^{\e-2}  e^{-\tau}e^{\frac{-\tilde\xi_0\eta_\rho-ia\tilde\xi_0}{\tau}} 
\\
\cdot\Big[1+O((-i\eta_\rho+a)^{-1}\tau)+O((-i\eta_\rho+a)^{-2}\tau^2 a\big)\Big]
d\tau,
\end{multline}
where  $\e>1$.

If   $\tilde C_n=\frac{\tilde C}{<\tilde C,\tilde C>^{\frac{1}{2}}}$  is  the normalized  wave  packet  then
\begin{equation}																		\label{eq:5.24}
\langle 0|N(C_n)|0\rangle
=\sum_{m'=-\infty}^\infty <\tilde C,\tilde C>^{-1}\int\limits_{-\infty}^\infty|\tilde C_1^- +\tilde C_2^-|^2d\eta_\rho,
\end{equation}
We compute  first  $\sum\limits_{m'=-\infty}^\infty <\tilde C,\tilde C>^{\frac{1}{2}}\int\limits_{0}^\infty|\tilde C_1^-|^2d\eta_\rho$.
It follows  from  
(\ref{eq:5.11})  and  (\ref{eq:5.22})  that  
\begin{align}																			\label{eq:5.25}
&\sum_{m'=-\infty}^\infty <\tilde C,\tilde C>^{-1}\int\limits_{0}^\infty|\tilde C_1^-|^2d\eta_\rho
\\ 
\nonumber
&\leq  C\int\limits_0^\infty\frac{a^{2\e -1}|\eta_\rho^2+a^2|^{\frac{1}{2}}}
{|-i\eta_\rho+a|^{2\e+2}}
d\eta_\rho
\leq C\int\limits_0^\infty\frac{d\eta_\rho}{|-i\eta_\rho+a|^2}\rw 0\ \ \mbox{when}\ \ a\rw\infty.
\end{align}
To compute  the term  containing  $|A_2|^2$  we need  the integration  by parts.

We have
\begin{equation}																		\label{eq:5.26}
\frac{d}{d\tau} e^{-\frac{i\tilde\xi_0}{\tau}(-i\eta_\rho +a)}=
\frac{i\tilde \xi_0(-i\eta_\rho+a)}{\tau^2}e^{-\frac{i\tilde\xi_0}{\tau}(-i\eta_\rho +a)}.
\end{equation}
Substituting  (\ref{eq:5.26}) in  (\ref{eq:5.23})  and integrating  by parts  in $\tau$,  we get
\begin{equation}															 			\label{eq:5.27}
A_2=\tilde\xi_0(-i\eta_\rho+a)^{-\e+1}\cdot O\Big(\frac{1}{-i\eta_\rho+a}\Big).
\end{equation}
Therefore
\begin{multline}																			\label{eq:5.28}
\sum_{m'=-\infty}^\infty <\tilde C,\tilde C>^{-1}\int\limits_{0}^\infty|\tilde C_2^-|^2d\eta_\rho
\leq C\int\limits_0^\infty
 \frac{a^{2\e-1}d\eta_\rho}
{
|-i\eta_\rho+a|^{2\e}(\eta_\rho^2+a^2)^{\frac{1}{4}}  
 }
\\
\leq C
\int\limits_0^\infty
\frac{d\eta_\rho}{|-i\eta_\rho+a|^2}\rw 0 
\ \ \mbox{when}\ \ a\rw \infty.
\end{multline}
Consider  now  $<\tilde C,\tilde C>^{-1}\sum\limits_{m'=-\infty}^\infty\int\limits_{-\infty}^0|\tilde C_1^-+\tilde C_2^-|^2d\eta_\rho$.

Making the changes of variables
\begin{equation}																		\label{eq:5.29}
\tau=at,\ \ \ \eta_\rho''=\frac{\eta_\rho}{a^2},
\end{equation}
we get  from (\ref{eq:5.19})
\begin{equation}																		\label{eq:5.30}
A_1(a^2\eta_\rho'',\tilde\xi_0,a)=a^{-\e-1}\int\limits_0^\infty e^{ia\eta_\rho'' \tau-\tau-\frac{i\tilde\xi_0 a}{\tau}}\tau^\e d\tau.
\end{equation}
Apply  the stationary phase  method   to the integral  
\begin{equation}																		\label{eq:5.31}
\int\limits_0^\infty e^{ia\gamma(\tau,\eta_\rho'')-\tau}\tau^\e d\tau,
\end{equation}
where 
\begin{equation}																		\label{eq:5.32}
\gamma(\tau,\eta_\rho)=\eta_\rho''\tau-\frac{\tilde\xi_0}{\tau}.
\end{equation}
The critical point  $\gamma_\tau(\tau_0,\eta_\rho'')=0$  is  $\eta_\rho'' +\frac{\tilde\xi_0}{\tau_0^2}=0$,  i.e.
$\tau_0=\frac{\sqrt{\tilde\xi_0}}{\sqrt {-\eta_\rho''}}$. 

Note that $\eta_\rho''<0$.  Since  
\begin{equation}																		\label{eq:5.33}
\gamma_\tau''(\tau,\eta_\rho'')=\frac{-2\tilde\xi_0}{\tau_0^3}=-2\tilde\xi_0^{-\frac{1}{2}}|\eta_\rho''|^{\frac{3}{2}}, 
\end{equation}
 we have
\begin{equation}																		\label{eq:5.34}
\int\limits_0^\infty e^{ia\gamma(\tau,\eta_\rho'')} e^{-\tau}\tau^\e dr=
\sqrt{\frac{\pi}{a}}
\frac{
e^{ia\gamma(\tau_0,\eta_\rho'')-i\frac{\pi}{4}}}
{\sqrt {2\tilde\xi_0^{-\frac{1}{2}}|\eta_\rho''|^{\frac{3}{2}}} }
e^{-\big(\frac{\tilde \xi_0}{|\eta_\rho''|}\big)^{\frac{1}{2}} }\Big(\frac{\tilde \xi_0}{|\eta_\rho''|}\Big)^{\frac{\e}{2}} +O\Big(\frac{1}{a}\Big).
\end{equation} 
Therefore,
\begin{align}																		\label{eq:5.35}
&\lim_{a\rw\infty}<\tilde C,\tilde C>^{-1}\sum_{m'=-\infty}^\infty\int\limits_{-\infty}^0 |\tilde C_1^-|^2d\eta_\rho
\\
\nonumber
=&\lim_{a\rw\infty} C'\int\limits_{-\infty}^0a^2 a^{2\e-1}
(a^4\eta_\rho''^2+a^2)^{\frac{1}{2}}a^{-2\e-2}\frac{1}{a}
\Bigg[\tilde\xi_0^{\frac{1}{2}}|\eta_\rho''|^{-\frac{3}{2}}
e^{-2\big(\frac{\tilde \xi_0}{|\eta_\rho''|}\big)^{\frac{1}{2}}} \Big(\frac{\tilde \xi_0}{|\eta_\rho''|}\Big)^\e +O\Big(\frac{1}{a}\Big)\Bigg]d\eta_\rho''
\\
\nonumber
= &C'\int\limits_{-\infty}^0   
 e^{-2\big(\frac{\tilde \xi_0}{|\eta_\rho''|}\big)^{\frac{1}{2}}} \Big(\frac{\tilde \xi_0}{|\eta_\rho''|}\Big)^{\e+\frac{1}{2}}
d\eta_\rho''
\end{align}
where   $C'$  in (\ref{eq:5.35})  and below means  various constants independent  of  $\tilde\xi_0$.  Changing  in  (\ref{eq:5.35})
$\eta_\rho''=\tilde\xi_0\tilde\eta_\rho$  we  get that  the right hand side  of  (\ref{eq:5.35})  is equal  to  $C'\tilde\xi_0$. 
Analogously consider  (\ref{eq:5.20}).
Note that 
 \begin{equation}																	\label{eq:5.36}
 f(\rho)=f(m+t)=-\frac{2 m}{m+t}+\frac{ m^2}{( m+t)^2}.
 \end{equation}
 Thus  $A_2(\eta_\rho,\tilde\xi_0,a)$  behaves  similarly to $A_1(\eta_\rho,\tilde\xi_0,a)$.
 
 Make changes  of variables  (\ref{eq:5.29})  in (\ref{eq:5.20}) and    compute  the stationary phase integral
 \begin{equation}												 					\label{eq:5.37}
 \int\limits_0^\infty e^{ia\eta_\rho''\tau -ia\frac{\hat\xi_0}{\tau}}e^{-\tau}\tau^{\e-2}\Big(\hat \xi_0+i(f-1)\big(\frac{\e}{a}\tau-\frac{\tau^2}{a}\big)\Big)d\tau.
 \end{equation}  
Thus  
\begin{align}																		\label{eq:5.38}
&<\tilde C,\tilde C>^{-1}\sum_{m'=-\infty}^\infty\int\limits_{-\infty}^0|C_2^-|^2d\eta_\rho
\\
\nonumber
&= C'\int\limits_{-\infty}^0
\frac{a^{2\e -1}}
{(a^4\eta_\rho''^2+a^2)^{\frac{1}{2}}
}
\cdot \frac{\tilde\xi_0^2}{a^{2(\e-1)}} \cdot \frac{1}{(\sqrt a)^2}
\tilde\xi_0^{\frac{1}{2}}|\eta_\rho''|^{-\frac{3}{2}}
\\
\nonumber
&\cdot
e^{-2\big(\frac{\tilde \xi_0}{|\eta_\rho''|}\big)^{\frac{1}{2}}} \Big(\frac{\tilde \xi_0}{|\eta_\rho''|}\Big)^{\e-2}
\Big(1+O\Big(\frac{1}{a}\Big)\Big) a^2
d\eta_\rho''
\\
\nonumber
&=C'\int\limits_{-\infty}^0
\frac{\tilde\xi_0^2}{|\eta_\rho''|^2}\Big(\frac{\xi_0}{|\eta_\rho''|}\Big)^{\frac{1}{2}}
e^{-2\big(\frac{\tilde \xi_0}{|\eta_\rho''|}\big)^{\frac{1}{2}}} \Big(\frac{\tilde \xi_0}{|\eta_\rho''|}\Big)^{\e-2}
\Big(1+O\Big(\frac{1}{a}\Big)\Big) a^2
d\eta_\rho''
\end{align}
Taking the limit  when  $a\rw\infty$   and changing  $\eta_\rho''=\tilde\xi_0\tilde\eta_\rho$,   we get
\begin{multline}																			\label{eq:5.39}
\lim_{a\rw\infty}\sum_{m'=-\infty}^\infty<\tilde C,\tilde C>^{-1}\int\limits_{-\infty}^0|\tilde C_2^-|^2d\tilde \eta_\rho
\\
=C'\tilde\xi_0\int\limits_{-\infty}^0  \frac{1}{|\tilde\eta_\rho''|^{2+\frac{1}{2}} }e^{-\frac{2}{|\tilde\eta_\rho|^{\frac{1}{2}}}}
\frac{1}{|\tilde\eta_\rho|^{\e-2}}d\tilde\eta_\rho=C'\tilde\xi_0.
\end{multline}
Analogously,  $\lim_{a\rw\infty}<\tilde C,\tilde C>^{-1}\int_{-\infty}^0 2\Re (C_1^-\overline{C_2^-})d\eta_\rho=C'\tilde\xi_0$.
Therefore
\begin{equation}                  																\label{eq:5.40}
\lim_{a\rw\infty}<\tilde C,\tilde C>^{-1}\int_{-\infty}^0|\hat \hat C_1^-+C_2^-|^2d\eta_\rho=C'\tilde\xi_0.
\end{equation}

Thus we have
\begin{equation}	
\nonumber																		
\lim_{a\rw \infty}\langle 0|N(\tilde C_n)|0\rangle  =I_1+I_2,
\end{equation}
where  $I_1$  is  the integral over $\int\limits_0^\infty$  and $I_2$  is the integral  over  $\int\limits_{-\infty}^0$.

We have  proved  that $I_1=0$  and  $I_2=C'\tilde\xi_0\neq 0$.
Note that  $I_1$  is the  portion  of the average  number  of particles that related to the Hawking radiation.

{\bf Remark 5.1}.
\ Consider  the Hawking radiation  for the extremal RN  black hole  when  the vacuum state is the Unruh type vacuum state as in \S 4.  Then
formula (\ref{eq:4.9})  holds with $C^{-+}(k)$  and  $C^{+-}(k)$  replaced by  $\tilde C^{-+}(k)$  and  $\tilde C^{+-}(k)$,  respectively.

 Normalizing $\tilde C(k)$  and taking the limit as  $a\rw  \infty$  we get  (see (\ref{eq:5.25})   and (\ref{eq:5.28})) 
that the contribution   of $\tilde C^{-+}(k)$  to the  Hawking  radiation  is zero.  Similarly computations show  that  the contribution  of 
  $\tilde C^{+-}(k)$ is also zero.  Thus
  \begin{equation}																		\label{eq:5.41}
  \lim_{a\rw\infty}  \langle \Psi|N(\tilde C_n)|\Psi\rangle=0.
  \end{equation}

\end{document}